\documentclass[a4paper,fleqn]{cas-sc}

\usepackage{bm}

\usepackage[authoryear,longnamesfirst]{natbib}

\def\tsc#1{\csdef{#1}{\textsc{\lowercase{#1}}\xspace}}
\tsc{HLG}
\tsc{VGR}

\begin{document}
\let\WriteBookmarks\relax
\def\floatpagepagefraction{1}
\def\textpagefraction{.001}

\shorttitle{Mixture model fitting using conditional models and modal Gibbs sampling}    

\shortauthors{Héctor L\'opez-G\'omez and Virgilio G\'omez-Rubio}  

\title [mode = title]{Mixture model fitting using conditional models and modal Gibbs sampling}  

\tnotemark[1] 

\tnotetext[1]{This work has been supported by grants PPIC-2014-001-P and SBPLY/17/180501/000491, funded by Consejer\'ia de
Educaci\'on, Cultura y Deportes (JCCM) and FEDER, grant MTM2016-77501-P,
funded by Ministerio de Econom\'ia y Competitividad and grant PID2022-136455NB-I00, funded by MCIN/AEI/10.13039/501100011033/FEDER.}

\author[1]{H. L\'opez-G\'omez}[orcid={0009-0007-7722-0676}]

\cormark[2]

\ead{Hector.Lopez4@alu.uclm.es}

\affiliation[1]{organization={Department of Mathematics, School of Industrial Engineering-Albacete},
            addressline={Avda. Espa\~{n}a s/n}, 
            city={Albacete},
            postcode={02071}, 
            country={Spain}}

\author[1]{V. G\'omez-Rubio}[orcid={0000-0002-4791-3072}]

\ead{Virgilio.Gomez@uclm.es}

\cortext[2]{Corresponding author}


\begin{abstract}
Mixture models are a convenient way of modeling data using a convex combination
of different parametric distributions.  A new algorithm based on Gibbs sampling
is used to approximate the posterior distribution of the auxiliary variables,
that assign each observation to a group in the mixture, without sampling any
other parameter in the model. In particular, the modes of an approximation to
the full conditional distributions of the parameters of the densities in the
mixture are computed using the Integrated Nested Laplace Approximation.  These
are plugged-in to the full conditional distribution of the auxiliary variables
to draw samples.  The posterior distributions of the remainder of the
parameters in the mixture are obtained by averaging over their conditional
posterior marginals on the auxiliary variables using Bayesian model averaging.

This approximation, 'modal' Gibbs sampling, reduces the computational burden in
the Gibbs sampling algorithm and provides very good estimates of the posterior
distribution of the auxiliary variables.  A simulation study supports the
validity of 'modal' Gibbs sampling and two examples on well-known datasets are
discussed using a mixture of Gaussian and Poisson distributions, respectively.
\end{abstract}



\maketitle

\section{Introduction}

Mixture models are a convenient way of describing data when these are thought to
come from different groups or components which are represented by
different distributions.  Some well known examples of this type of data include
the velocities of galaxies \citep{CarlinChib:1995,Chib:1995}, waiting and
eruption times of the Old Faithful geyser in the Yellowstone National Park
\citep{AzzaliniBowman:1990} and the number of fetal movements in lambs 
\citep{LerouxPuterman:1992,Chib:1996}.  In all these cases, the data
generating process is made of different components with different parametric
densities and the observed data does not include information about the
component to which observations belong. For this reason, mixture models are
often represented using latent auxiliary variables to indicate to which
component each observation belongs and the parametric distribution of each
component.

In this paper a new approach to fitting mixture models is developed by focusing
on estimating the posterior distribution of the auxiliary variables first. Once
this posterior distribution has been obtained, the posterior marginal of the
remainder of the parameters in the model is obtained by Bayesian model
averaging over conditional  posterior distributions given the values of the
auxiliary variables. This is possible because, given the value of the
auxiliary variables, a mixture model with exchangeable priors becomes a model
with different independent components and the posterior marginals of their
parameters can easily be obtained.

Hence, model fitting will be split in two steps. First of all, the posterior
distribution of the latent auxiliary variables will be estimated.  This will be
done by means of what we have called 'modal' Gibbs sampling (MGS) because some
of the parameters will not be sampled but the modes of their full conditional
distributions will be used instead.  Secondly, conditional models on these
auxiliary variables will be fit by considering values of the auxiliary
variables with non-negligible posterior probability.  These conditional models
will be fit with the integrated nested Laplace approximation
\citep[INLA,][]{Rueetal:2009} during MGS and the resulting
conditional marginals will be combined using Bayesian model averaging
\citep[BMA,][]{Hoetingetal:1999,Bivandetal:2014}.  As conditional models are fit with
INLA, it is possible to consider a wide range of distributions for the mixture
components and the use of non-conjugate priors for their parameters.
Furthermore, this approach seems to be less likely to suffer from label
switching and it allows for the computation of measures for the selection of
the optimal number of components, such as the marginal likelihood.

The analysis of mixture models using Bayesian inference has been considered by
several authors \citep[see, for example,][for a
discussion]{Chib:1995,Cappeetal:2002,Marinetal:2005}.  More recently,
\citet{GomezRubioRue:2018} describe how to combine INLA within the
Metropolis-Hastings algorithm to fit mixture. However, in this paper a compltely
new approach is considered based on estimating the posterior distribution of
the auxiliary variables first.



The paper starts with an introduction to mixture models and Bayesian inference
in Section~\ref{sec:mixmod}. Next, the new proposal to fit mixture models is
described in Section~\ref{sec:MGS}.  A simulation study is carried out in
Section~\ref{sec:simulation}.  Several examples are developed in
Section~\ref{sec:examples}. Finally, the developments and main results of this
paper are discussed in Section~\ref{sec:discussion}.

\section{Bayesian Inference on Mixture Models}
\label{sec:mixmod}


Given a set of $n$ observations $\mathbf{y} = (y_1,\ldots,y_n)$, a mixture model with
$K$ components or groups is usually defined as follows:

$$
y_i \sim \sum_{j=1}^K w_j f_j(y_i|\theta_j),\ i=1,\ldots,n
$$
\noindent
Here, $f_j(\cdot|\theta_j)$ is the distribution of group or component $j$ in
the mixture, which is defined by a set of parameters $\theta_j$. Parameters
$w_j$ indicate the weight of each component in the mixture and these are taken to sum up to 1,
i.e., $\sum_{j=1}^K w_j = 1$. Weights depend on the number of observations
that belong to each component and the vector of weights $\{w_j\}_{j=1}^K$ 
will be denoted by $\mathbf{w}$.
The ensemble of parameters $\{\theta_j\}_{j=1}^K$ will be denoted by vector
$\bm \theta$ and a single element of it by $\theta_{t}\in\bm\theta$.

Typically, the form of distributions $f_j(\cdot|\theta_j)$, and their
associated parameters $\theta_j$, will depend on the actual problem. For
example, for continuous observations, $f_j(\cdot|\theta_j)$ can be a Normal
distribution with parameters $\theta_j = (\mu_j, \tau_j)$, where $\mu_j$ is the
mean and $\tau_j$ the precision. When the observations are count data,
distributions $f_j(\cdot|\theta_j)$ could well be Poisson with different means,
i.e., $\theta_j = \lambda_j$.


An alternative formulation of the model, very useful for Bayesian inference
with MCMC, is to express the mixture by including auxiliary variables
$\{z_i\}_{i=1}^n$ to indicate to which component observation $i$ belongs.
These auxiliary variables are defined so that the probability of observation
$i$ belonging to component $j$, conditional on $\mathbf{w}$, is:

\begin{equation}
\pi(z_i = j|\mathbf{w}) =  w_j,\ j=1,\ldots,K
\end{equation}
\noindent
This means that, given $z_i = j$, the distribution of $y_i$ is 
$f_{z_i}(\cdot|\theta_{z_i})$. The ensemble of auxiliary variables
$(z_1, \ldots , z_n)$ will be denoted by $\mathbf{z}$.

%
%
%

We have followed \citet{Marinetal:2005} in order to provide a brief summary of
Bayesian inference for mixture models below.  First of all, observations $y_i$
are independent given $z_i$ and $\theta_{z_i}$, hence:

\begin{equation}
\pi(\mathbf{y}|\mathbf{z}, \bm \theta) =
  \prod_{i=1}^n f_{z_i}(y_i|\theta_{z_i})
\end{equation}

Note also that the distribution on $\mathbf{z}$ 
given weights $\mathbf{w}$ is defined as:

$$
\pi(\mathbf{z}|\mathbf{w}) = \prod_{j=1}^K w_j^{n_j}
$$
\noindent
where $n_j$ is the number of observations assigned to component $j$.

Finally, in order to complete our Bayesian formulation of the model, priors on
$\bm \theta$ and $\mathbf{w}$ must be defined.  For convenience, conjugate
priors are often used as this will make inference using MCMC easier. However, as
we will see in Section~\ref{sec:MGS}, sampling on $\bm \theta$ and $\mathbf{w}$
will not be done in practice.  The prior on $\bm \theta$ will depend on the
distributions in the mixture. For example, for a mixture of Normal
distributions, means $\{\mu_j\}_{j=1}^K$ can be assigned a vague Normal prior
(e.g., centered at zero with a small precision) whilst precisions
$\{\tau_j\}_{i=1}^K$ can be assigned vague Gamma priors.  For a mixture of
Poisson distributions, means $\{\mu_j\}_{j=1}^K$ can be assigned a vague Gamma
prior. However, given that model fitting will be done with INLA using numerical
integration there is a wider choice of priors available.

Weights $\mathbf{w}$ can be assigned a Dirichlet prior with parameters
$(\alpha_1,\ldots, \alpha_K)$.  A convenient vague prior on $\mathbf{w}$ is a
Dirichlet with parameters $\alpha_1 = \ldots = \alpha_K < d/2$, where
$d$ is the number of parameters in the densities of the mixture \citep{RousseauMengersen:2011}.

It is worth noting that the prior on $\bm \theta$ can be used to include some
identifiability constraints in the model \citep[see, e.g.,][]{CarlinChib:1995}.
For example, in a mixture model with two components defined by Normal
distributions with means $\mu_1$ and $\mu_2$, respectively, the prior may be
informative so that $\mu_1 < \mu_2$, or include a constraint to make sure that
$\mu_1 < \mu_2$. This will make the two components fully identifiable. Hence,
the prior in this case could be proportional to $\pi(\mu_1)\pi(\mu_2)I(\mu_1 <
\mu_2)$, where $I(\cdot)$ is the indicator function.


\section{'Modal' Gibbs sampling for mixture models}
\label{sec:MGS}

It should be noted that, given $\mathbf{z}$, the observations are assigned to a
particular group in the mixture and the model can be expressed as a model with
several likelihoods, which is easier to fit.  In addition, the posterior
marginals of the parameters in the mixture model can be obtained by averaging
over the posterior distribution of $\mathbf{z}$ as follows: 

$$
\pi(\theta_{t}|\mathbf{y}) = \sum_{z\in\mathcal{Z}} \pi(\theta_{t}|\mathbf{y}, \mathbf{z} = z)
\pi(\mathbf{z} = z|\mathbf{y}),\ \theta_{t}\in\bm\theta
$$
\noindent
Here, $\mathcal{Z}$ is the parameter space of the auxiliary variables,
which is the $n$-dimensional Cartesian product of $\{1,\ldots, K\}$.

Note that it is not necessary to explore $\mathcal{Z}$ exhaustively and
exploring $\mathcal{Z}^* \subset \mathcal{Z}$ so that

$$
\sum_{z\in\mathcal{Z} \backslash \mathcal{Z}^*} \pi(z|\mathbf{y}) \simeq 0
$$
\noindent
will be enough in practice. Note also that this estimation method only relies
on $\mathbf{z}$ and conditional models fit with INLA, and the simulated values
of $\bm\theta$ and $\mathbf{w}$ are not necessary.

For this reason, the new algorithm to fit mixture models proposed now 
can be regarded as a collapsed Gibbs sampling \citep[see ][]{Liu:1994,MillerHarrison:2018} in the
sense that some of the parameters are integrated out in the full conditionals
and not all parameters in the model are sampled.  During Gibbs sampling
only new values of $\mathbf{z}$ will be sampled given the (conditional) modes
of $\mathbf{w}$ and $\bm \theta$, i.e., samples from $\mathbf{w}$ and $\bm
\theta$ are replaced by the modes of their respective full conditional
distributions or an approximation to them.

The full conditional distribution of $\mathbf{w}$ is a Dirichlet distribution
with parameters equal to $(\alpha_1 + n_1, \ldots, \alpha_K + n_K)$, which has
a mode at $((\alpha_1 + n_1 - 1)/n^{*}, \ldots, (\alpha_K + n_K)/n^{*})$, with
$n^{*} =n + \sum_{i=1}^K \alpha_i
- K$. If for some reason the mode has any negative value, it will be replaced
by its mean.
The modes of the conditional
distributions of $\bm \theta$ used in the Gibbs sampling algorithm can be
obtained in several ways. In this particular case, they will be provided by
numerical approximation with the Integrated Nested Laplace Approximation
\citep[INLA, see][]{Rueetal:2009}.


Hence, INLA can be used to obtain accurate approximations of conditional
marginals $\pi(\theta_{t}|\mathbf{y},\mathbf{z})$, denoted by
$\tilde\pi(\theta_{t}|\mathbf{y},\mathbf{z})$.  INLA is implemented as a
package for the R statistical software \citep{R:2018} called R-INLA. This
package implements a number of latent effects and prior distributions, and
allows for an easy model fitting and visualization of the output
\citep[see][for a recent review]{Rueetal:2017}. 

INLA (and R-INLA) can handle different likelihoods, so conditional models on
$\mathbf{z}$ can be easily fit.  Furthermore, an approximation to the
marginal likelihood is also provided $\pi(\mathbf{y})$, so that model selection
\citep{GomezRubioetal:2017} and Bayesian model averaging
\citep{Bivandetal:2014} can be carried out. This approximation is denoted by
$\tilde{\pi}(\mathbf{y})$, and \citet{HubinStorvik:2016b} and
\citet{GomezRubioRue:2018} have shown that this approximation is very accurate
on a wide range of models.

Modal Gibbs sampling will take, at each iteration, the mode of the
full conditional of $\mathbf{w}$ and the modes of the distributions
$\tilde\pi(\theta_{t}|\mathbf{y},\mathbf{z})$. Note that this implies
the following two assumptions:

\begin{enumerate}

\item Conditional distribution
$\tilde\pi(\theta_{t}|\mathbf{y},\mathbf{z})$ is a good approximation to
full conditional  distribution
$\tilde\pi(\theta_{t}|\mathbf{y},\mathbf{z},\bm\theta_{-t})$, i.e.,
$\theta_{t}$ and $\bm\theta_{-t}$ are (approximately) conditionally
independent given $\mathbf{y}$ and $\mathbf{z}$. 

\item The approximation to the mode provided by $\tilde\pi(\theta_{t}|\mathbf{y},\mathbf{z})$ is also accurate.
\end{enumerate}

These two main assumptions are discussed in
Appendix~\ref{app:MGS} for mixtures of Gaussian and Poisson distributions. In
general, these assumptions will be valid when exchangeable priors are used, and also when
the number of parameters in
$\theta_j,j=1,\ldots,K$ is equal to one (e.g., a Gaussian with
known precision or a Poisson distribution)
or the number of data allocated to each component is large.
Figure~\ref{fig:EMINLA} summarizes modal Gibbs sampling using INLA
to obtain the approximations to the conditional distributions needed.

Note that MGS can also be regarded as a variation of the EM algorithm
\citep{Dempsteretal:1977} applied to a regularized likelihood (i.e., Maximum A
Posteriori, MAP, estimation), because of the prior, in which the E-step is
replaced by sampling the values of $\mathbf{z}$. For this reason MGS i also
close to Monte Carlo EM algorithms \citep{LevineCasella:2001}.

\begin{figure}t
\begin{footnotesize}
\begin{enumerate}

\item  Assign initial values to $\mathbf{z}$: $\mathbf{z}^{(0)}$.

\item For $l=1, 2, \ldots$, repeat:

   \begin{enumerate}

  \item Fit model (conditional on $\mathbf{z}^{(l-1)}$) with INLA to approximate
conditional marginals $\pi(\theta_{t}|\mathbf{y}, \mathbf{z}^{(l-1)}),\ \theta_{t} \in \bm\theta$.

  \item Obtain (conditional) modes of $\mathbf{w}$ and
$\bm \theta$: $\mathbf{\hat{w}}^{(l)}$ and $\bm{\hat{\theta}}^{(l)}$.

   \item  Sample $z_i^{(l)}$ from $ \pi(\cdot|\mathbf{y}, \mathbf{\hat{w}}^{(l)}, \bm{\hat{\theta}}^{(l)})$:

  $$
  \pi(z_i = j|\mathbf{y},\mathbf{\hat{w}}^{(l)}, \bm{\hat{\theta}}^{(l)}) \propto \hat{w}_j f_j(y_i|\hat{\theta}_j);\ i=1,\ldots,n;\ j=1,\ldots,K
  $$

   \end{enumerate}

\item Estimate the posterior marginals of the model parameters as:

$$
\tilde\pi(\theta_{t}|\mathbf{y}) = \sum_{z\in\mathcal{Z^*}} \tilde\pi(\theta_{t}|\mathbf{y}, \mathbf{z} = z)
\hat\pi(\mathbf{z} = z|\mathbf{y}),\ \theta_{t}\in\bm\theta
$$

\end{enumerate}
\end{footnotesize}
\caption{'Modal' Gibbs sampling algorithm to fit mixture models with INLA.}
\label{fig:EMINLA}
\end{figure}

\subsection{Accuracy of 'modal' Gibbs sampling}
\label{subsec:accuracyMGS}

The critical point in this approach is the reliability of 'modal' Gibbs
sampling to provide estimates $\hat\pi_G(\mathbf{z} | \mathbf{y})$ 
which are a good approximations to $\pi(\mathbf{z} | \mathbf{y})$.
Similarly as stated in \citet{GomezRubioRue:2018} for the Metropolis-Hastings
algorithm, because the approximations in 'modal' Gibbs sampling are
deterministic, then the algorithm will converge to $a$ stationary distribution.
The question is whether this distribution is close enough to $\pi(\mathbf{z} |
\mathbf{y})$.  This essentially means that all assignments $z\in\mathcal{Z}$
with non-negligible probability are explored  and that their probabilities are
estimated with accuracy. 

Let us denote by $\mathcal{Z}^* \subseteq \mathcal{Z}$ the subset of all
possible assignments explored by 'modal' Gibbs sampling. Because of the
structure of 'modal' Gibbs sampling in which different assignments are
simulated by conditioning on the modes of the parameters of the distributions
of the components, we believe that the parameter space $\mathcal{Z}$ is
conveniently explored as this mimics the data generating process. All the
assignments left out are likely to have a very small probability of occurring
under the data generating process \citep[see,][for a similar
discussion]{Porteousetal:2008}.  Furthermore, the simulation study in
Section~\ref{sec:simulation} supports that the stationary distribution obtained
with MGS is close enough to $\pi(\mathbf{z} |\mathbf{y})$ to obtain good
estimates of $\pi(\theta_t|\mathbf{y})$ and $\pi(\mathbf{y})$.

Alternatively, samples from MGS can be assigned weights that will ensure that
their posterior distribution is close to the posterior. This follows, for
example, the argument given in \citet{Cappeetal:2004} for Population Monte
Carlo in the sense that instead of providing an proof of convergence, good
approximations to the posterior probabilities are obtained by re-weighting
the observations with weights proportional to

$$
\frac{\pi(\mathbf{y}|\mathbf{z})\pi(\mathbf{z})}{\hat\pi_G(\mathbf{z} | \mathbf{y})}.
$$

 Alternatively, assuming that only assignments in $\mathcal{Z}^*$ have a
non-negligible probability, the posterior probability of
$\pi(\mathbf{z} = z^* |\mathbf{y}), z^*\in\mathcal{Z}^*$, can be approximated as

\begin{equation}
\pi(\mathbf{z}= z^*|\mathbf{y}) \simeq
\hat\pi_I(\mathbf{z} = z^*|\mathbf{y}) =
\frac{\tilde\pi(\mathbf{y}|\mathbf{z} = z^*)\pi(\mathbf{z} = z^*)} {\sum_{z\in\mathcal{Z}^*}
\tilde\pi(\mathbf{y}|\mathbf{z} = z)\pi(\mathbf{z} = z)}
\label{eq:postz}
\end{equation}

These probabilities can be used as weights for the unique values of
$\mathbf{z}$ obtained from MGS to compute summary statistics on the posterior.
Note that this approximation is easy and fast to compute from the conditional
models fit during MGS.

Hence, this can be computed and compared to the probabilities obtained with the
samples from MGS, $\hat\pi_G(\mathbf{z|\mathbf{y}})$. A
simple diagnostics is to plot $\hat\pi_I(\mathbf{z} = z^*|\mathbf{y})$
against 
$\hat\pi_G(\mathbf{z} = z^*|\mathbf{y})$ for all $z^*\in\mathcal{Z}^*$. If
both estimates of the posterior probabilities do not match, then there is
reason to suspect that 'modal' Gibbs sampling has not explored the parameter
space $\mathcal{Z}$ conveniently. In this case, $\hat\pi_I(\mathbf{z}|\mathbf{y})$ should be preferred as an estimate of the posterior probabilities.


As a general criterion to assess convergence of the chain generated by MGS is
to plot the series of conditional marginal log-likelihoods approximations
$\tilde\pi(\mathbf{y}|\mathbf{z})$. These values
should increase during the burn-in period and then reach some stationary state
if MGS has converged to a region of high posterior probability close to the
posterior mode.  Hence, this plot can also be used as a diagnostics.

All the previous algorithms will provide  approximations to the posterior of
$\mathbf{z}$ and the conditional posterior marginal distributions
$\pi(\theta_{t}|\mathbf{z}, \mathbf{y})$ for the model parameters
$\theta_{t}\in\bm\theta$. The
actual posterior marginals of the model parameters can be approximated as

\begin{equation}
\pi(\theta_{t}|\mathbf{y}) \simeq \sum_{z\in\mathcal{Z}} \tilde\pi(\theta_{t}|\mathbf{z} = z^*, \mathbf{y})
  \hat\pi(\mathbf{z} = z^* | \mathbf{y}) 
\end{equation}
\noindent
where weights $\hat\pi(\mathbf{z} = z^* | \mathbf{y})$ are taken from 
estimates 
$\hat\pi_G(\mathbf{z}|\mathbf{y}) $ or $\hat\pi_I(\mathbf{z|\mathbf{y}})$, as
discussed above.

The full conditional distribution of $\mathbf{w}$ is the same as in
Section~\ref{sec:mixmod}. i.e., a Dirichlet distribution. Similarly,
the posterior distribution of $\mathbf{w}$ 
can be approximated as

$$
\pi(\mathbf{w}|\mathbf{y}) \simeq \sum_{z\in\mathcal{Z}} \pi(\mathbf{w}|\mathbf{z} = z, \mathbf{y})
  \hat\pi(\mathbf{z} = z| \mathbf{y})
$$
\noindent
Note that the previous distribution is a mixture of Dirichlet distributions,
so it can be computed very efficiently.

\citet{GriffithsSteyvers:2004} have proposed a similar algorithm to assign
documents to a mixture of topics. However, in their case they are able to
derive the full conditional of $z_i$ on $z_{-i}$ and $\mathbf{y}$, so there is
no need to deal with parameters $\bm\theta$ during Gibbs sampling.  Once they
have obtained the posterior distribution on the auxiliary variables
$\mathbf{z}$ they are able to compute the posterior distribution on the
remainder of the model parameters by averaging over $\mathbf{z}$ using its
posterior distribution.
\citet{Porteousetal:2008} follow a similar approach by they develop a fast
algorithm by arguing that, for a given document, posterior probabilities will
be concentrated on a number of topics. Hence, they first explore the set of 
topics fast and then refine the estimated posterior probabilities.

\subsection{Selecting the number of components}
\label{sec:components}

Selecting the optimal number of components in the mixture is an important
problem in mixture models. From a Bayesian perspective, this implies computing
the posterior probabilities of the number of components  over a range of
values. The problem of choosing the number of components in the mixture can be
regarded as a model selection problem as well, which can be tackled by
computing the marginal likelihood of different mixture models.  See
\citet{FruhwirthSchnatter:2006} for a general discussion on determining
marginal likelihoods for Bayesian finite mixture models in detail.

First of all, the marginal likelihood of a given model can also be expressed
as

$$
\pi(\mathbf{y}) = \sum_{z\in\mathcal{Z}} \pi(\mathbf{y}|\mathbf{z} = z) \pi(\mathbf{z} = z),
$$
\noindent
which means that it can be approximated as

\begin{equation}
\pi(\mathbf{y}) \simeq \hat\pi_S(\mathbf{y}) = \sum_{z^*\in\mathcal{Z}^*} \hat\pi(\mathbf{y}|\mathbf{z} = z^*) \pi(\mathbf{z} = z^*).
\label{eq:piI}
\end{equation}
\noindent
Note that  $\pi(\mathbf{z})$  is \citep[see, for example,][]{GriffithsSteyvers:2004}:

$$
\pi(\mathbf{z}) = \int_{\mathcal{W}} \pi(\mathbf{z}|\mathbf{w}) \pi(\mathbf{w}) d\mathbf{w} =
\frac{\Gamma(\sum_{i=1}^K \alpha_i)}{\prod_{i=1}^K \Gamma(\alpha_i)}
\frac{\prod_{i=1}^K \Gamma(n_i + \alpha_i)}{\Gamma(n + \sum_{i=1}^K \alpha_i)}
$$
\noindent
Here, $\Gamma(\cdot)$ is the Gamma function and $\mathcal{W}$ is the parametric
space of the weights vector $\mathbf{w}$.

Given models $\mathcal{M}_1, \ldots,\mathcal{M}_p$, we can identify each model
by its number of components (i.e., $\mathcal{M}_1$ has one component,
$\mathcal{M}_2$ has two components, and son on). Hence, a
first approach for model selection is to approximate their
marginal likelihoods $\{\hat\pi(\mathbf{y}|\mathcal{M}_k) \}_{k=1}^p$
and choose the model with the highest one.

Alternatively, each model can be assigned a prior probability
$\pi(\mathcal{M}_k),\ k=1,\ldots,p$, so that the posterior probability of each
model can be computed as:

\begin{equation}
\pi(\mathcal{M}_k|\mathbf{y}) = \frac{\pi(\mathbf{y}|\mathcal{M}_k)\pi(\mathcal{M}_k)}{\sum_{k=1}^p \pi(\mathbf{y}|\mathcal{M}_k)\pi(\mathcal{M}_k)}
\label{eq:postprob}
\end{equation}

\noindent
Here, $\pi(\mathbf{y}|\mathcal{M}_k)$ is the marginal likelihood of model
$\mathcal{M}_k$.
The marginal likelihood $\pi(\mathbf{y}|\mathcal{M}_k)$ of a given model
$\mathcal{M}_k$ can be approximated as above to approximate its posterior 
probability.

Alternatively, \citet{Chib:1995} and \citet{ChibJeliazkov:2001} describe
several approaches to compute the marginal likelihood of a model using MCMC.
They note that the marginal likelihood is the scaling constant when
writing the posterior distribution using Bayes' rule. For mixture models, this
can be written by conditioning on $\mathbf{z}$ as 

\begin{equation}
\pi(\mathbf{y}) = \frac{\pi(\mathbf{y} | \mathbf{z})\pi(\mathbf{z})}{\pi(\mathbf{z} | \mathbf{y})}
\label{eq:ppmodel}
\end{equation}
\noindent
Note that this holds for any $\mathbf{z}$ but, as pointed out by
\citet{Chib:1995} and \citet{ChibJeliazkov:2001}, values with a high posterior
probability (or density, if $\mathbf{z}$ is a continuous random variable) are
preferred since $\pi(\mathbf{z} | \mathbf{y})$ will be away from zero. In this
case, posterior mode $ \mathbf{z^m}$ could be used to compute the
log-marginal likelihood as:

\begin{equation}
\log(\pi(\mathbf{y})) = \log(\pi(\mathbf{y} | \mathbf{z^m})) + \log(\pi(\mathbf{z^m}))
  - \log(\pi(\mathbf{z^m} | \mathbf{y}))
\end{equation}
\noindent
Note that the first and second terms in the right hand side are the conditional
marginal likelihood (for which an accurate approximation is provided by INLA)
and the prior evaluated at the posterior mode of $\mathbf{z}$, respectively.
The last term needs to be obtained from the MCMC output. In practice, the
following approximation will be used:

\begin{equation}
\log(\pi(\mathbf{y})) \simeq \log(\tilde\pi(\mathbf{y} | \mathbf{z^m})) + \log(\pi(\mathbf{z^m}))
  - \log(\hat\pi(\mathbf{z^m} | \mathbf{y}))
\label{eq:logpiy}
\end{equation}

Note that the last term in Equation~(\ref{eq:logpiy}) can be approximated in
different ways, that will lead to different approximations to
$\pi(\mathbf{y})$. If it is the estimate provided by modal Gibbs sampling,
$\hat\pi_G(\mathbf{z}|\mathbf{y})$, then
the approximation to the marginal likelihood will be denoted by
$\hat\pi_G(\mathbf{y})$.  If the last term is computed using the conditional
marginal likelihoods provided by INLA as in Equation~(\ref{eq:postz}),
$\hat\pi_I(\mathbf{z} | \mathbf{y})$, the
approximation will be denoted by $\hat\pi_I(\mathbf{y})$.

Computing these three different estimates of the marginal likelihood can be a
way to perform an assessment of whether the parametric space $\mathcal{Z}$ has
been conveniently explored and the posterior probabilities of its elements
estimated with accuracy. In particular, $\hat\pi_G(\mathbf{y})$ will be very
sensitive to this issue and large differences with $\hat\pi_S(\mathbf{y})$ and
$\hat\pi_I(\mathbf{y})$ will indicate that the posterior probabilities provided
by modal Gibbs sampling may not be accurate and that samples must be reweighted
accordingly.

Estimates of the posterior probability of a given model, as defined in
Equation~(\ref{eq:postprob}), can be obtained with any of the three estimates of
the marginal likelihood and they will be denoted similarly.  For example,
$\hat\pi_G(\mathcal{M}_i | \mathbf{y}),\ i=1,\ldots,p$ are the estimates of
the posterior probabilities of the different models based on estimates
$\{\hat\pi_G(\mathbf{y}|\mathcal{M}_k)\}_{k=1}^p$ of their marginal likelihoods.

\section{Simulation Study}
\label{sec:simulation}

In order to assess the performance of 'modal' Gibbs sampling, a simulation
study is presented in this Section.  The aim of this study is twofold, as the
interest is in estimating both the posterior marginal distributions of the
model parameters and the actual number of components in the mixture.

We have followed \citet{Nasserinejadetal:2017} and simulated four different
datasets. The first one is made of 150 observations from a Gaussian
distribution with mean 1 and standard deviation 0.25.  This represents the case
of a dataset with a single group. The other three datasets have been simulated
using different degrees of separation between the different components. In all
cases, we have three components (i.e., $K=3$) with means $\mu_1=1$, $\mu_2=2$
and $\mu_3 = 3$ and we have simulated 50 observations from each
component.  For a high separation the standard deviation of all components is
$0.25$, for a mild separation the standard
deviations are all equal to $0.4$ and for a low separation the standard
deviations are all equal to $0.7$.

\begin{figure}
\centering
\includegraphics[width=6cm]{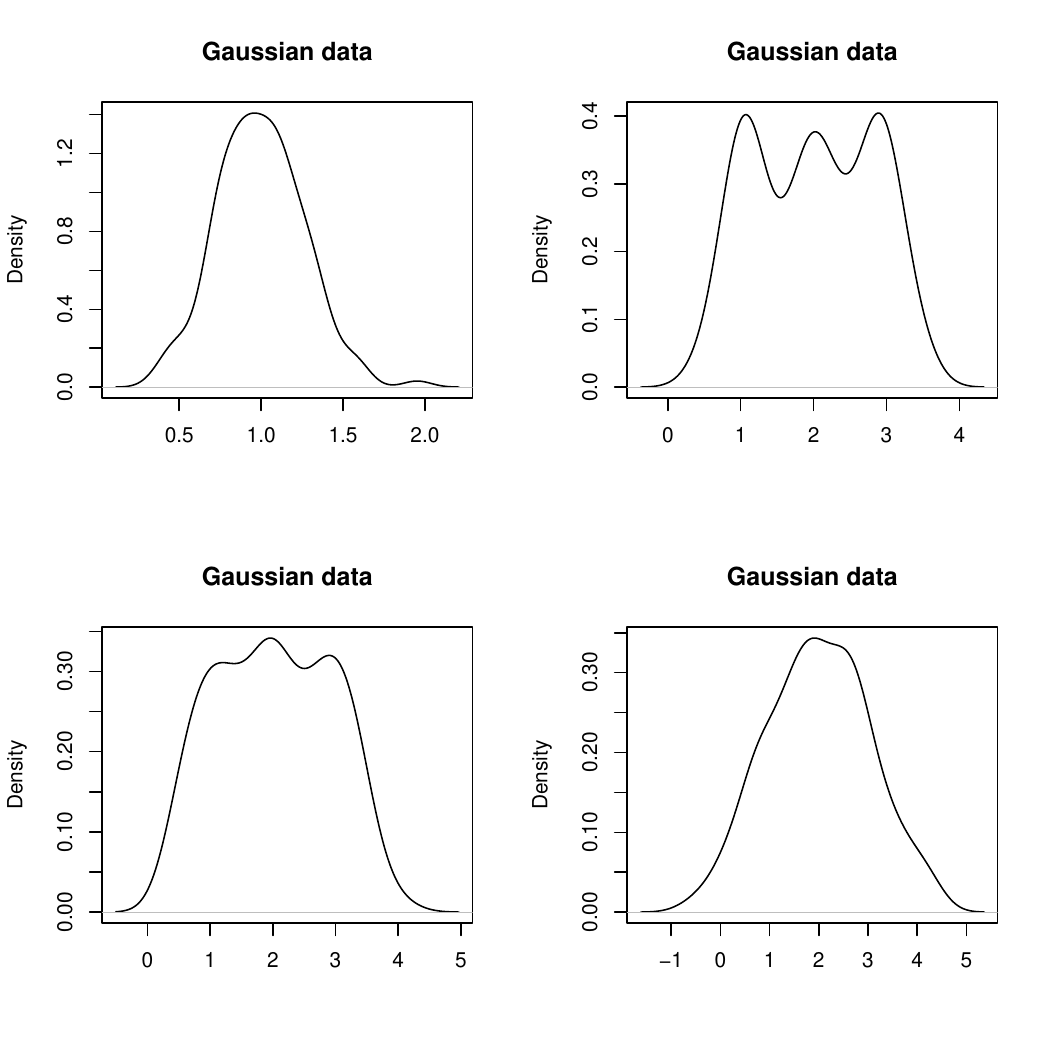}
\caption{Simulated data from a single Gaussian distribution (top-left), and
high (top-right), mild (bottom-left) and low (bottom-right) separation between the groups.}
\label{fig:sim-data}
\end{figure}

Mixture models with up to 5 components will be fit to the data to assess model
choice and the estimation of the model parameters.  A density estimate of the
simulated data can be found in Figure~\ref{fig:sim-data}.

In principle, all $K$ components will not be assumed to have the same variances
and the groups will be represented by $K$ Gaussian distributions with different
means and variances.  Furthermore, the results obtained with MGS will be
compared to those obtained with MCMC  with the JAGS
software \citep{rjags:2016}. For this, a burn-in of 200 iterations plus other
10000 iterations have been used, of which only one every ten has been kept, so
that a final 1000 samples have been used for inference.  The means of the
Gaussian distributions in the mixture have been ordered in increasing order at
every step of the sampling process with JAGS in order to reclass the assignment
to the mixture components to reduce label switching. 

In order to define the full Bayesian model, we have used a Gaussian prior on
the means with zero mean and precision 1/1000. For the precisions, we have used
a Gamma distribution with parameters 0.5 and 0.5. The parameter of the
Dirichlet prior will be a vector of the appropriate length with values equal to
0.4 following \citet{Nasserinejadetal:2017}. Also, when considering model
selection, all models have been assigned the same probability a priori.

\begin{table}[pos=h]
\centering
\begin{footnotesize}
\caption{Results of simulation study using a mixture of Gaussian distributions. 
Symbol $^*$ means that 
model $\mathcal{M}_1$ has entirely been computed with INLA as it has a single component.}
\label{tab:sim-gau}
\setlength{\tabcolsep}{0.5em}

\begin{tabular}{|c|c|c|c|c|}
\multicolumn{5}{c}{\textbf{Dataset with 1 group}}\\
Model & $\log(\hat\pi_S(\mathbf{y}))$ & $\log(\hat\pi_G(\mathbf{y}))$ & $\hat\pi_S(\mathcal{M}_j|\mathbf{y})$ & $\hat\pi_G(\mathcal{M}_j|\mathbf{y})$\\
\hline
$\mathcal{M}_1$$^*$ &  -28.55 & -28.55 &  0.90 & 0.74\\
$\mathcal{M}_2$ & -30.90 & -32.51  & 0.09 & 0.01\\
$\mathcal{M}_3$ & -33.44 & -29.81  & 0.01 & 0.21\\
$\mathcal{M}_4$ & -35.48 & -31.82  &  0.00 & 0.03\\
$\mathcal{M}_5$ & -37.51 & -33.36  &  0.00 & 0.01\\
\end{tabular}

\begin{tabular}{|c|c|c|c|c|}
\multicolumn{5}{c}{\textbf{Dataset with 3 groups (high separation)}}\\
Model & $\log(\hat\pi_S(\mathbf{y}))$ & $\log(\hat\pi_G(\mathbf{y}))$ &  $\hat\pi_S(\mathcal{M}_j|\mathbf{y})$ & $\hat\pi_G(\mathcal{M}_j|\mathbf{y})$\\
\hline
$\mathcal{M}_1$$^*$ &  -192.25 & -192.25 &  0.18 & 0.02\\
$\mathcal{M}_2$ & -190.76 & -188.18 &   0.82 & 0.98\\
$\mathcal{M}_3$ & -199.34 & -195.59 &  0.00  & 0.00\\
$\mathcal{M}_4$ & -208.96 & -285.75 &  0.00 & 0.00\\
$\mathcal{M}_5$ & -219.39 & -310.01 &  0.00 & 0.00\\
\end{tabular}

\begin{tabular}{|c|c|c|c|c|}
\multicolumn{5}{c}{\textbf{Dataset with 3 groups (mild separation)}}\\
Model & $\log(\hat\pi_S(\mathbf{y}))$ & $\log(\hat\pi_G(\mathbf{y}))$ & $\hat\pi_S(\mathcal{M}_j|\mathbf{y})$ & $\hat\pi_G(\mathcal{M}_j|\mathbf{y})$\\
\hline
$\mathcal{M}_1$$^*$ &  -208.17 & -208.17 &  0.99  & 0.23\\
$\mathcal{M}_2$ & -227.67 & -250.97 & 0.00 & 0.00\\
$\mathcal{M}_3$ & -213.15 & -206.94 &  0.01 & 0.77\\
$\mathcal{M}_4$ & -215.18 & -351.36 &  0.00 & 0.00\\
$\mathcal{M}_5$ & -239.11 & -403.82 &  0.00 & 0.00\\
\end{tabular}

\begin{tabular}{|c|c|c|c|c|}
\multicolumn{5}{c}{\textbf{Dataset with 4 groups (low separation)}}\\
Model & $\log(\hat\pi_S(\mathbf{y}))$ & $\log(\hat\pi_G(\mathbf{y}))$ & $\hat\pi_S(\mathcal{M}_j|\mathbf{y})$ & $\hat\pi_G(\mathcal{M}_j|\mathbf{y})$\\
\hline
$\mathcal{M}_1$$^*$ &  -226.69 & -226.69 & 0.83 & 0.16\\
$\mathcal{M}_2$ & -228.36 & -225.04 &  0.16 & 0.84\\
$\mathcal{M}_3$ & -230.98 & -312.42 &  0.01 & 0.00\\
$\mathcal{M}_4$ & -251.00 & -389.16 &  0.00 & 0.00\\
$\mathcal{M}_5$ & -250.37 & -452.30 &  0.00 & 0.00\\
\end{tabular}

\end{footnotesize}
\end{table}

\begin{table}[pos=h]
\centering
\caption{Results of simulation study using a mixture of Gaussian distributions.
Symbol $^*$ means that
model $\mathcal{M}_1$ has entirely been computed with INLA as it has a single component. Parameter estimates are summarized using
posterior mean and standard deviation (between parentheses).}
\label{tab:sim-gau2}
\setlength{\tabcolsep}{0.5em}
\begin{scriptsize}

\begin{tabular}{|c|c|c|c|c|c|c|}
\multicolumn{7}{c}{\textbf{Dataset with 1 group}}\\
Model  & Parameter & Comp. 1 & Comp. 2 & Comp. 3 & Comp. 4 & Comp. 5\\ 
\hline
$\mathcal{M}_1$$^*$ &  $\mu$ & 1.00 (0.02) & -- & -- & -- & -- \\
 & $\tau$ & 13.01 (1.50)& -- & -- & -- & -- \\
$\mathcal{M}_2$ & $\mu$ & 0.98 (0.03) & 1.23 (1.68) & -- & -- & -- \\
 & $\tau$ & 13.76 (2.26) & 9036.53 (33876.40) & -- & -- & -- \\
 & $\mathbf{w}$ & 0.89 (0.18) & 0.11 (0.18) & -- & -- & -- \\
$\mathcal{M}_3$ & $\mu$ & 0.44 (20.25) & 0.99 (0.02) &
   1.11 (4.31) & -- & -- \\
& $\tau$ & 3.89e+04 (1.20+e05) & 13.57 (2.19) & 1.78e+05 (1.11e+05) & -- & --\\
& $\mathbf{w}$ & 0.06 (0.12) & 0.87 (0.15) & 0.06 (0.08) & -- & -- \\
$\mathcal{M}_4$ & $\mu$ &  0.18 (24.93) & 0.66 (15.72) & 1.00 (0.03) & 0.99 (7.88) & -- \\
& $\tau$ & 1.20 e+05 (2.76e+05) & 2.65e+04 (8.75e+04) & 13.26 (2.32) & 1.30e+04 (4.86e+04) & --\\
 & $\mathbf{w}$ & 0.01 (0.03) & 0.10 (0.16) & 0.85 (0.18) & 0.04 (0.05) & -- \\
$\mathcal{M}_5$ &  $\mu$ & 0.12 (25.87) & 0.27 (23.16) & 0.75 (12.29) & 0.99 (0.05) & 0.87 (12.09)\\
 & $\tau$ & 1.03e+05 (2.62e+05) & 6.15e+04 (18.98e+04) & 1.97e+04 (6.77e+04) & 30.29 (286.55) & 2.07e+04 (8.11e+04\\
 & $\mathbf{w}$ & 0.01 (0.10) & 0.02 (0.04) & 0.14 (0.20) & 0.80 (0.22) & 0.03 (0.04)\\
\end{tabular}

\begin{tabular}{|c|c|c|c|c|c|c|}
\multicolumn{7}{c}{\textbf{Dataset with 3 groups (high separation)}}\\
Model & Parameter & Comp. 1 & Comp. 2 & Comp. 3 & Comp. 4 & Comp. 5\\
\hline
$\mathcal{M}_1$$^*$ & $\mu$ & 2.01 (0.07) & -- & -- & -- & -- \\
 & $\tau$ & 1.47 (1.17)& -- & -- & -- & -- \\
$\mathcal{M}_2$ &  $\mu$ & 1.04 (0.05) & 2.47 (0.06) & -- & -- & --\\
 & $\tau$ & 13.58 (3.44) & 3.16 (0.53) & -- & -- & --\\
 & $\mathbf{w}$ & 0.34 (0.02) & 0.68 (0.02) & -- & -- & -- \\
$\mathcal{M}_3$ & $\mu$ & 1.03 (0.04) & 2.01 (0.06) & 2.93 (0.05) & -- & --\\
 & $\tau$ & 14.43 (3.33) & 11.45 (3.74) & 11.92 (2.87) & -- & --\\
 & $\mathbf{w}$ & 0.33 (0.01) & 0.32 (0.03) & 0.34 (0.02) & -- & --\\
$\mathcal{M}_4$ & $\mu$ &  1.04 (1.24) & 1.81 (0.45) & 2.14 (0.48) & 2.86 (0.18) & --\\
 & $\tau$ & 3932.60 (1.88e+05) & 86.60 (89.50) & 121.48 (114.62) & 16.71 (200.41) & -- \\
 & $\mathbf{w}$ & 0.25 (0.10) & 0.23 (0.14) & 0.20 (0.17) & 0.32 (0.11) & -- \\
$\mathcal{M}_5$ & $\mu$ & 0.61 (20.51) & 1.53 (4.03) & 1.78 (0.63) & 2.11 (0.90) & 2.87 (0.16) \\
 & $\tau$ & 6.93e+04 (1.40e+05) & 1.36e+04 (4.33e+04) & 99.62 (101.70) & 176.25 (1821.21) & 6.60 (26.05)\\
 & $\mathbf{w}$ & 0.12 (0.12) & 0.15 (0.13) & 0.25 (0.18) & 0.18 (0.16) & 0.30 (0.10)\\
\end{tabular}

\end{scriptsize}
\end{table}

\begin{table}[pos=h]
\centering
\caption{Results of simulation study using a mixture of Gaussian distributions.
Symbol $^*$ means that
model $\mathcal{M}_1$ has entirely been computed with INLA as it has a single component. Parameter estimates are summarized using
posterior mean and standard deviation (between parentheses).}
\label{tab:sim-gau2b}
\setlength{\tabcolsep}{0.5em}
\begin{scriptsize}
\begin{tabular}{|c|c|c|c|c|c|c|}
\multicolumn{7}{c}{\textbf{Dataset with 3 groups (mild separation)}}\\
Model & Parameter & Comp. 1 & Comp. 2 & Comp. 3 & Comp. 4 & Comp. 5\\
\hline
$\mathcal{M}_1$$^*$ &  $\mu$ & 2.02 (0.08) & -- & -- & -- & -- \\
 & $\tau$ & 1.19 (1.14)& -- & -- & -- & -- \\
 & $\mathbf{w}$ & -- & -- & -- & -- & -- \\
$\mathcal{M}_2$ &  $\mu$ & 1.23 (0.18) & 2.70 (0.19) & -- & -- & --\\
 & $\tau$ & 4.08 (1.68) & 3.20 ( 1.24) & -- & -- & --\\
 & $\mathbf{w}$ & 0.46 (0.12) & 0.53 (0.12) & -- & -- & --\\
$\mathcal{M}_3$ &  $\mu$ & 1.11 (0.23) & 1.95 (0.49) & 2.87 (0.21) & -- & --\\
 & $\tau$ & 66.90 (465.82) & 98.95 (97.64) &  4.30 (1.97) & -- & --\\
 & $\mathbf{w}$ & 0.33 (0.13) & 0.29 (0.17) & 0.38 (0.15) & -- & --\\
$\mathcal{M}_4$ &  $\mu$ & 1.00 (3.82) & 1.61 (0.48) & 2.12 (0.80) & 2.88 (0.25) & --\\
 & $\tau$ & 1.53e+04 (4.40e+04) & 82.06 (90.90) & 106.72 (103.93) & 20.24 (225.89 & --\\
 & $\mathbf{w}$ & 0.19 (0.13) & 0.25 (0.16) & 0.23 (0.17) & 0.33 (0.17) & --\\
$\mathcal{M}_5$ & $\mu$ & 0.76 (13.53) & 1.40 (3.20) & 1.76 (0.82) & 2.24 (0.97) & 2.93 (0.23)\\
 & $\tau$ & 1.71e+04 (6.26e+04) & 6501.62 (2.76e+04) & 133.93 (126.65) & 127.95 (123.47) & 11.97 (129.90)\\
 & $\mathbf{w}$ & 0.10 (0.11) & 0.18 (0.16) & 0.24 (0.18) & 0.20 (0.18) & 0.27 (0.14)\\
\end{tabular}
\begin{tabular}{|c|c|c|c|c|c|c|}
\multicolumn{7}{c}{\textbf{Dataset with 3 groups (low separation)}}\\
Model & Parameter & Comp. 1 & Comp. 2 & Comp. 3 & Comp. 4 & Comp. 5\\
\hline
$\mathcal{M}_1$$^*$ & $\mu$ & 2.00 (0.085) & -- & -- & -- & --\\
 & $\tau$ & 0.93 (0.11) & -- & -- & -- & -- \\
 & $\mathbf{w}$ & -- & -- & -- & -- & -- \\
$\mathcal{M}_2$ & $\mu$ & 1.37 (0.60) & 2.46 (0.44) & -- & -- & --\\
 & $\tau$ & 72.35 (86.52) & 18.47 (185.24) & -- & -- & --\\
 & $\mathbf{w}$ & 0.43 (0.30) & 0.57 (0.30) & -- & -- & -- \\
$\mathcal{M}_3$ &  $\mu$ & 1.10 (0.99) & 1.93(0.47) & 2.75 (0.48) & -- & --\\
 & $\tau$ & 137.31 (130.34) & 13.99 (75.30) & 49.24 (416.48) & -- & --\\
 & $\mathbf{w}$ & 0.25 (0.19) & 040 (0.25) & 0.35 (0.23) & -- & --\\
$\mathcal{M}_4$ & $\mu$ & 0.88 (4.47) & 1.57 (1.08) & 2.15 (0.43) & 2.85 (0.60) & --\\
 & $\tau$ & 1.03e+05 (8.49e+05) & 131.41 (126.88) & 21.67 (235.47) & 94.44 (95.75) & --\\
 & $\mathbf{w}$ & 0.14 (0.14) & 0.23 (0.19) & 0.39 (0.26) & 0.24 (0.20) & --\\
$\mathcal{M}_5$ & $\mu$ & 0.61 (11.31) & 1.25 (2.75) & 1.75 (0.68) & 2.21 (0.52) & 2.88 (0.69) \\
 & $\tau$ & 2.18e+05 (1.31e+05) & 5780.65 (2.70e+04) & 88.69 (90.47) & 99.79 (96.50) & 72.06 (83.43)\\
 & $\mathbf{w}$ & 0.07 (0.09) & 0.14 (0.13) & 0.22 (0.18) & 0.35 (0.24) & 0.21 (0.19)\\
\end{tabular}
\end{scriptsize}
\end{table}

%

Table~\ref{tab:sim-gau} shows the estimates of the marginal likelihoods of the
simulation study for the Gaussian data.  Table~\ref{tab:sim-gau2} and
Table~\ref{tab:sim-gau2b} summarize point estimates of the parameters of the
different components in the mixture using MGS. In all scenarios, the model with
the right number of components provides very good point estimates of the
model parameters.

Model selection using the marginal likelihood only seems to be reliable for the
case of 1 component and, partially, for the case of three components and high
separations (where the model with 2 components is selected).  This is in line
with the findings reported by \citet{Nasserinejadetal:2017}, that also
report the difficulty of selecting the right number of components for
data with mild and low separation. Alternatively,
model selection can be done in practice looking at the estimates of
$\mathbf{w}$ and $\bm\tau$ because groups which are superflous have a small
weight estimate and a extremly high estimate of the precision. By using this
criterion models with more components that needed can be identified in all
cases. Furthermore, for the model with the right number of components we have
assessed that the posterior marginals are very similar using the posterior
probabilities $\hat\pi_G(\mathbf{z}|\mathbf{y})$ and
$\hat\pi_I(\mathbf{z}|\mathbf{y})$ as weights for the conditional marginals
$\tilde\pi(\theta_t|\mathbf{y}, \mathbf{z})$.

Regarding our MCMC implementation with JAGS, we have observed that the models
with three components are not correctly identified, and MCMC converges to a
model with 2 components. However, when the parameters of the Dirichlet prior
are set to 2, then the identification of the components in the data improves.
This may be due to the fact that this ensures that at least one observation is
assigned to each group. Although this may not be ideal for models with many
components \citep[as discussed in][]{Nasserinejadetal:2017}, in our case it
helps to find the right number of components.  Note that we have not observed
these problems when fitting the mixture models with MGS and INLA.

\section{Examples}
\label{sec:examples}

\subsection{Gaussian mixture models: galaxy data}

\citet{CarlinChib:1995} and \citet{Chib:1995} have studied the velocities of 82
galaxies in the Corona Borealis region using mixture models.  They have fit
mixture models with different components to try to estimate the optimal number
of components in the mixture using different methods.  \citet{CarlinChib:1995}
compute the posterior probability of each model using MCMC methods, while
\citet{Chib:1995} focuses on computing the marginal likelihood to derive Bayes
factors for model choice. Mixture models with a different number of
components will be fit to this dataset in order to obtain the parameter
estimates and the marginal likelihood of each model, so that model choice can
be performed.

\begin{figure}
\centering
\includegraphics[width=5cm]{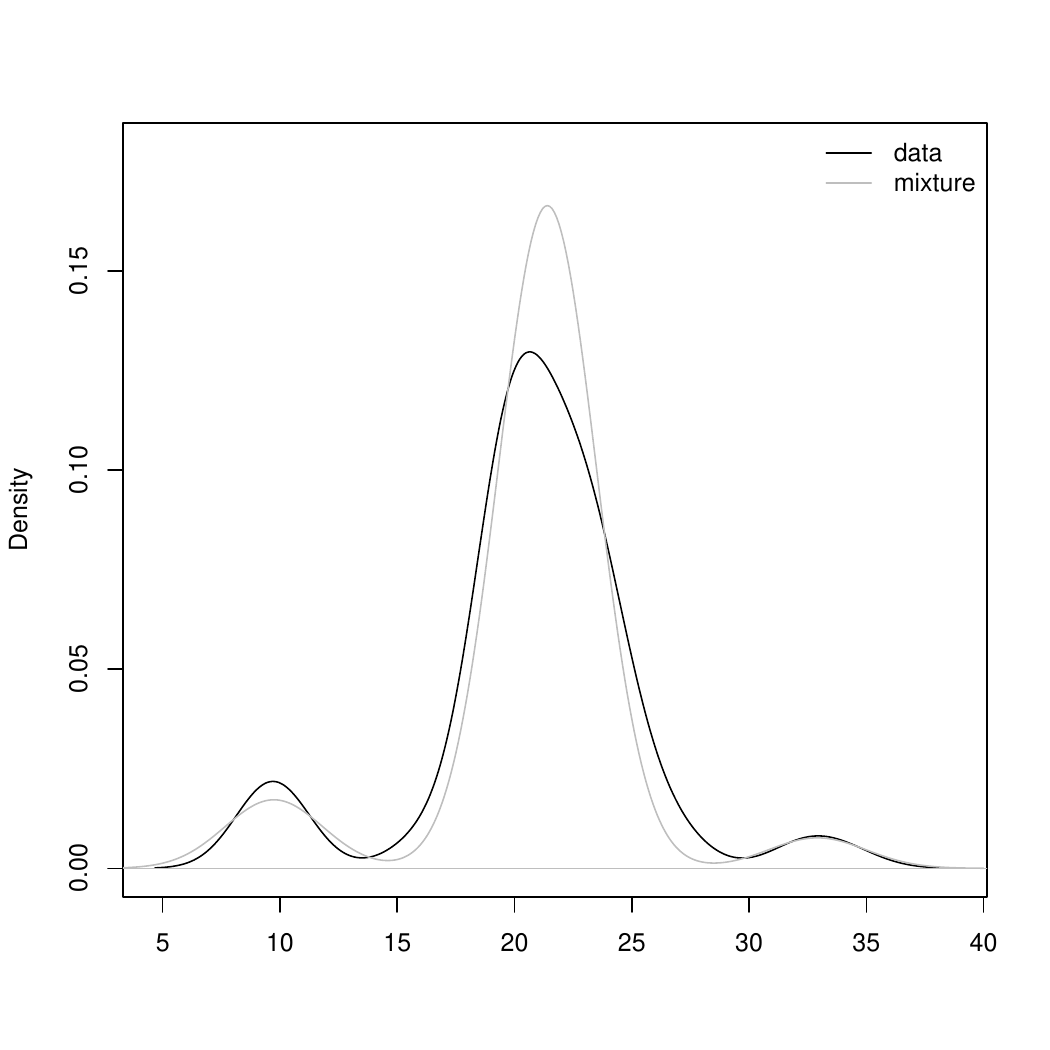}
\includegraphics[width=10.5cm]{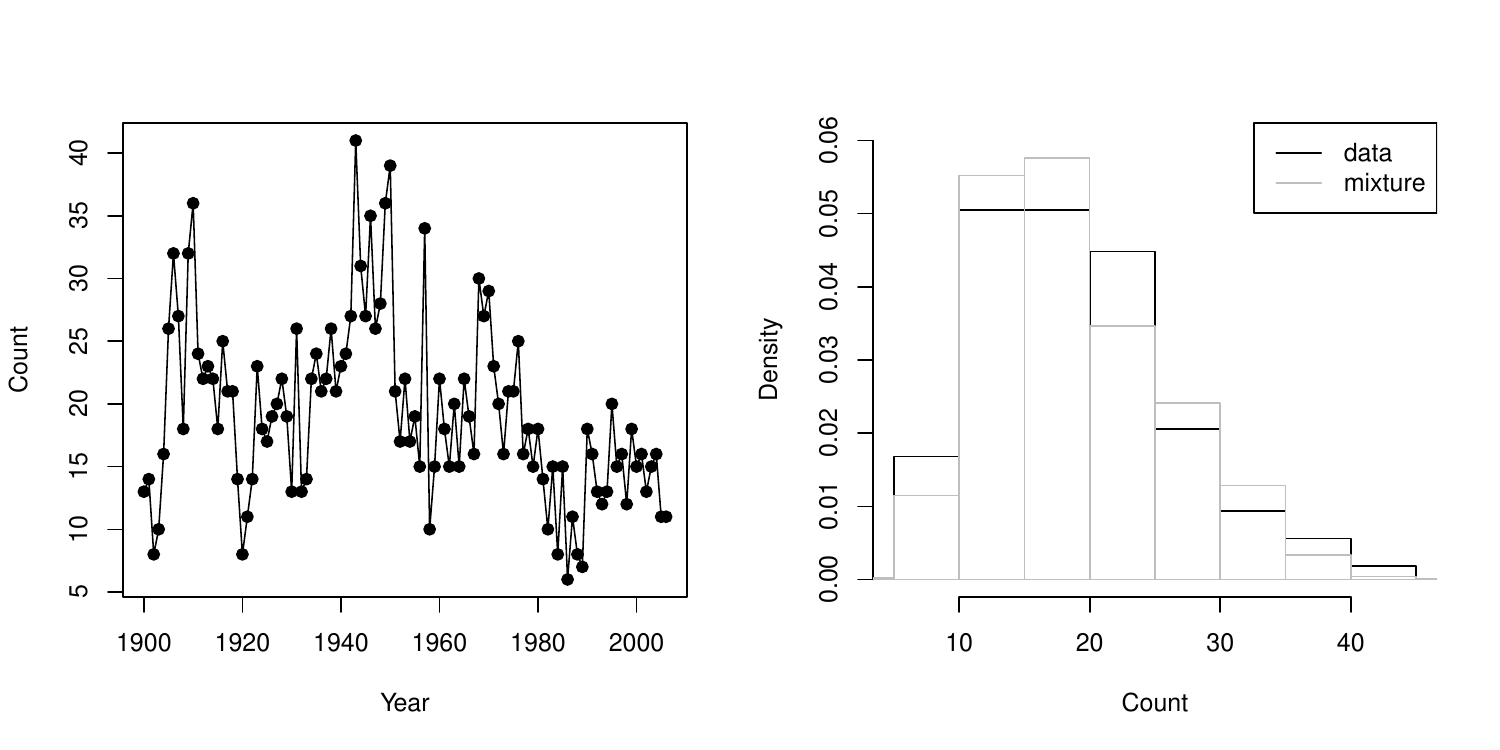}
\caption{Density estimate and estimated mixture with 3 components on the speeds of 82 galaxies in the
Corona Borealis region (left plot), number of yearly major earthquakes from 1900 to 2006 (center plot) and histogram of the earthquake data with the estimated mixture with two components (right plot).}
\label{fig:data}
\end{figure}

Figure~\ref{fig:data} (left plot) shows a kernel density estimate of the data,
where three groups seem to appear.  Hence, it makes sense to fit a mixture
model to the data. In order to consider several models, mixture models with up
to 4 components have been fit to the data.

In this case, all models are a mixture of Gaussian distributions. We have
considered a vague Gaussian prior (with zero mean and precision 0.001) for the
means, and a Gamma (with parameters 0.5 and 0.5) for the precisions.
Furthermore, all precisions have been considered to be equal, i.e., all
components have the same precision \citep[as in][]{CarlinChib:1995}. For all
models, simulations included 200 burn-in iterations followed by other 1000
iterations for inference, obtained after thinning (by keeping one in ten iterations).

\begin{table}[pos=h]
\centering
\begin{scriptsize}
\caption{Summary of results for the different mixture models fit on the Galaxies dataset. 
Symbol $^*$ means that
model $\mathcal{M}_1$ has entirely been computed with INLA as it has a single component. 
Parameter estimates are summarized using
posterior mean and standard deviation (between parentheses).}
\label{tab:galaxies}
\setlength{\tabcolsep}{0.5em}
\begin{tabular}{|c|c|c|c|c|c|c|c|c|c|}
Model & $\log(\hat\pi_S(\mathbf{y}))$ & $\log(\hat\pi_G(\mathbf{y}))$ & $\hat\pi_S(\mathcal{M}_j|\mathbf{y})$ & $\hat\pi_G(\mathcal{M}_j|\mathbf{y})$&
  \multicolumn{5}{c}{Parameter estimates}\\
 & $\log(\tilde\pi_M(\mathbf{y}))$& & & & Param. & Comp. 1 & Comp. 2 & Comp. 3 & Comp. 4\\
\hline
$\mathcal{M}_1$ &  -248.09$^*$ &  -248.09$^*$ & 0.00 & 0.00 & $\bm\mu$ &20.82 (0.51) & -- & -- & -- \\
 & & & & & $\tau$ & \multicolumn{4}{c|}{0.05 (0.007) }\\
$\mathcal{M}_2$ & -243.88 & -243.93 &  0.00& 0.00 & $\bm\mu$ &9.85 (1.27) &  21.87 (0.36) & -- & -- \\
 & & & & & $\mathbf{w}$ & 0.09 (0.011) & 0.91 (0.011) & --& -- \\
 & & & & & $\tau$ & \multicolumn{4}{c|}{0.11 (0.017) }\\
$\mathcal{M}_3$ & -232.06 & -233.15 & 1.00 & 1.00 & $\bm\mu$ & 9.74 (0.82) & 21.40 (0.25) & 
   32.91 (1.26) & -- \\
  & & & & &  $\mathbf{w}$ & 0.09 (0.002) & 0.87 (0.004) & 0.04 (0.003) & -- \\
 & & & & & $\tau$ & \multicolumn{4}{c|}{0.23 (0.036) }\\
$\mathcal{M}_4$ & -239.41 & -241.55 & 0.00 &0.00 &  $\bm\mu$&  9.71 (0.53) & 20.07 (0.31) &  23.48 (0.40) &  33.02 (0.81)\\
 & & & & & $\mathbf{w}$&0.09 (0.001) & 0.53 (0.055) & 0.34 (0.055) & 0.04 (0.000) \\ 
 & & & & & $\tau$ & \multicolumn{4}{c|}{0.53 (0.120) }\\
\end{tabular}
\end{scriptsize}
\end{table}

Table~\ref{tab:galaxies} summarizes the results obtained after fitting the
different mixture models. The values of the marginal likelihood with all three
methods are very similar, as well as the posterior probabilities. In addition,
\citet{Chib:1995} report the marginal likelihoods of the model with two and
three components and these values are very similar to the ones shown in
Table~\ref{tab:galaxies}.  The small discrepancies may be related to the choice
of the priors.  Furthermore,  \citet{CarlinChib:1995} report point estimates of
the model parameters for the models with 3 and 4 components, which are very
similar to the ones in Table~\ref{tab:galaxies} as well.  In this case, the
results favor the mixture model with three components. Hence, MGS 
seems to be a valid approach to fit Gaussian mixture models given that
similar results to other published bibliography have been obtained.

In order to provide an example with non-conjugate priors, we have considered a
half-Cauchy prior on the standard deviation of the Gaussian distributions in
the mixture. This follows \citet{Gelman:2006} to provide a less informative
prior on the scale parameters and illustrates that the approach described in
this paper is flexible enough to handle non-conjugate priors with ease as the
only modification required is to set this prior in the R code and INLA will
handle it.  Similarly, we have fit models with an improper uniform prior on the
standard deviations, but the use of improper priors is discouraged as mixture
models may not be identifiable in this case.  In both cases the results are
very similar to those provided in Table~\ref{tab:galaxies} (see
Appendix~\ref{app:galaxies}).

Finally, \citet{RichardsonGreen:1997} have also analyzed this dataset using
Reversible Jump MCMC methods, which allows for the computation of the posterior
probability of the number of components.  Their findings suggest a higher number
of components, between 5 and 7.

\subsection{Non-Gaussian mixture models: earthquake data}

The next example will consider distributions of non-Gaussian mixtures to
analyze 
the yearly number of major earthquakes (magnitude 7 or greater)
from 1990 to 2006.  Figure~\ref{fig:data} (center and right
plots) shows the actual dataset and a histogram of the yearly counts.  This
dataset has been analyzed in \citet{Zucchinietal:2016} where it is suggested
that this dataset could be analyzed using a mixture model given its
overdispersion.  Given that now the response variable represents counts, it
makes sense to use a mixture of Poisson distributions.


For this reason, different mixture models with up to four components have been
fit to the data.  The prior on the log-mean of each group is a vague Gaussian
distribution, with zero mean and precision equal to 0.001. Note that this prior
is used, instead of a typical conjugate Gamma prior on the mean
\citep{Viallefontetal:2002}, because the prior is on the intercept of the
linear predictor, which
results in a log-Gaussian prior on the mean.

The results are summarized in Table~\ref{tab:earthquakes}. MGS seems to provide
a different estimates of the marginal likelihood to the other two methods.
However, the point estimates of the model parameters are quite reasonable.
According to the point estimates of $\mathbf{w}$, the model with four
components seem to have a negligible component, while the model with 3
components seem to
have three solid ones.

\citet{Zucchinietal:2016} report results where the chosen model is the one with
three components. However, they use a different parameterization of the model,
as well as different priors on the mean of the groups.  This may be the reason
why model choice differs. Point estimates of the means are very similar to
those reported in \citet{Zucchinietal:2016} though.

\begin{table}[pos=h]
\caption{Summary of results for the different mixture models fit on the
earthquakes dataset.  Symbol $^*$ means that model $\mathcal{M}_1$ has entirely
been computed with INLA as it has a single component.  Parameter estimates are
summarized using posterior mean and standard deviation (between parentheses).}
\label{tab:earthquakes}
\centering
\begin{scriptsize}
\setlength{\tabcolsep}{0.5em}
\begin{tabular}{|c|c|c|c|c|c|c|c|c|c|c|}
Model & $\log(\hat\pi_S(\mathbf{y}))$ & $\log(\hat\pi_G(\mathbf{y}))$ & $\hat\pi_S(\mathcal{M}_j|\mathbf{y})$ &
$\hat\pi_G(\mathcal{M}_j|\mathbf{y})$ &
  \multicolumn{4}{c}{Parameter estimates}\\
 & $\log(\hat\pi_M(\mathbf{y}))$  & & $\hat\pi_M(\mathcal{M}_j|\mathbf{y})$ &  & Parameter & Comp. 1 & Comp. 2 & Comp. 3 & Comp. 4\\
\hline
$\mathcal{M}_1$$^*$ & -399.18 &-399.18 & 0.00 & 1.00 & $\bm\mu$ & 2.96 (0.02) & -- & --  & -- \\
$\mathcal{M}_2$ & -388.74 & -398.51 & 1.00 & 0.00 & $\bm\mu$ & 15.75  (0.73) & 26.94 (1.50)  & -- & -- \\
 & & & & & $\mathbf{w}$ & 0.67 (0.067) & 0.33 (0.066) & -- & -- \\
$\mathcal{M}_3$ & -398.55 & -433.63 & 0.00 & 0.00 &$\bm\mu$ & 12.28 (1.61) & 19.39 (1.43) & 31.05 (2.54) & --\\
 & & & & &  $\mathbf{w}$ & 0.25 (0.116) & 0.59 (0.105) & 0.16 (0.056) & -- \\
$\mathcal{M}_4$  & -403.73 & -468.61 &  0.00 & 0.00 &$\bm\mu$ & 0.00 (0.00) &  12.58 (1.74)  & 19.68 (1.56) & 30.95 (2.47)  \\
& & & & &  $\mathbf{w}$ & 0.01 (0.023) & 0.28 (0.129) & 0.57 (0.115) & 0.15 (0.057) \\
\end{tabular}
\end{scriptsize}
\end{table}

\section{Discussion}
\label{sec:discussion}

This paper introduces a novel approach to fit mixture models based on obtaining
first the posterior distribution of some latent auxiliary variables using a
variation of Gibbs sampling called 'modal' Gibbs sampling (MGS). Then, the
conditional (on the latent auxiliary variables) posterior marginals of the
remainder of the parameters in the mixture model are obtained with the
Integrated Nested Laplace Approximation (INLA). Finally, the posterior
marginals of these parameters are obtained by Bayesian model averaging of their
conditional posterior marginals. Because INLA is used to fit the conditional
models, the distributions of the components can be taken from a wide range of
distributions, conjugate priors are not required and label switching does not
seem to be a problem.

The MGS algorithm employed in model fitting only draws samples from
the auxiliary variables. For the other parameters, instead of sampling from
their full conditional distributions their respective modes are used.  The
modes are available as part of the output provided by INLA at every step of the
Gibbs sampling algorithm when a conditional model on the current value of the
auxiliary variables is fit. This simplifies the model fitting algorithm as
well as reduces computing time as actual sampling is reduced. Given the results
obtained in the simulation study, this simple approximation to the full
conditional of the model parameters appears to be enough to obtain a good
estimate of the posterior distribution of the indicator variables.  This
algorithm resembles collapsed Gibbs sampling \citep{Liu:1994} because some of
the model parameters are integrated out in the full conditionals.
MGS is also similar to Population Monte Carlo and Maximization A Posteriori
algorithms.

This novel approach also provides a simple way to compute the posterior
probabilities of the number of components, by exploiting the marginal
likelihood provided by INLA and the output from MGS. The simulation study
carried out supports that this is a valid inferential framework for mixture
models.  The results in the simulation study and the case studies suggest that
the estimates provided are very accurate.

Furthermore, if MGS is thought to provide biased estimates of the posterior
probabilities of the auxiliary variables, then the posterior probabilities can
be reweighted or recomputed with the estimates of the marginal likelihoods
provided by INLA, similarly as in Population Monte Carlo. This is similar to
what \citet{Porteousetal:2008} have done in the context of the classification
of documents by topic. They develop a fast Gibbs sampling algorithm for latent
Dirichlet allocation that explores the set of possible topics very quickly and
then refine the posterior probabilities because they argue that the probability
mass will be concentrated in a small number of topics.

The idea of obtaining estimates of the model parameter by averaging over the
indicator variables has been considered by several authors. See, for example,
\citet[][page 523]{BDA3:2013}, for a general overview, and
\citet{GriffithsSteyvers:2004}, for an early application of this idea on the
determination of topics covered in documents.  However, instead of averaging
over single values of the model parameters, this new approach performs Bayesian
model averaging on the (conditional) posterior distributions of the model
parameters obtained with INLA. Hence, no samples from the model parameters are
required, which simplifies model fitting. 

Finally, we believe that this approach can be applied to mixture models in a
variety of areas. For example, it can be used to fit cure rate models
\citep{BerksonGage:1952} in survival analysis, which are expressed as a mixture
model with two components.
 Similarly, mixture models could be fit to
classify areas which follow a shared spatio-temporal risk pattern or which
depart from this shared pattern in disease mapping \citep[see, for example,][]{Abellanetal:2008}.
Extending MGS to consider a multivarite response is also straightforward
as multivarite models can be easily be fit with INLA.



%

\printcredits

\bibliographystyle{cas-model2-names}

\bibliography{local,extra,mixtures}

\newpage

\appendix

%
%
%
%
%
%
%
%
%
%
%
%

\section{Approximation of 'modal' Gibbs sampling}
\label{app:MGS}

The 'modal' Gibbs sampling (MGS) algorithm is an approximation to
Gibbs sampling (GS) and, for this reason, some insight on when it is a
reasonable approximation needs to be provided. 
First of all, it should be noted that the approximation made by MGS can be
split in two steps:

\begin{enumerate}

\item Full conditional distribution  $\pi(\theta_{t}|\mathbf{y}, \mathbf{z}, \bm\theta_{-t})$ is replaced by $\pi(\theta_{t}|\mathbf{y}, \mathbf{z})$, i.e., it is assumed that $\theta_{t}$ and $\bm\theta_{-t}$ are independent given $\mathbf{z}$ and $\mathbf{y}$.

\item Sampling from $\pi(\theta_{t}|\mathbf{y}, \mathbf{z})$ is replaced
by its mode $\theta^*_{t}$, i.e., this can be regarded as approximating the previous distribution by a Dirac distribution about $\theta^*_{t}$.

\end{enumerate}

If these two steps provide accurate approximations to sampling from the full
conditional of $\theta_{t}$ then MGS will converge to a stationary distribution
that will be close to the actual one obtained with GS, i.e., the posterior
distribution of the model parameters. Otherwise, MGS will converge to a
different stationary distribution  and that this one is, in principle, unknown.
Note that if MGS is thought not to converge to the required distribution then
posterior probabilities of $\mathbf{z}$ can be computed or the samples
reweighted using the marginal likelihood provided by INLA as explained in
Section~\ref{subsec:accuracyMGS}.

Regarding the validity of step 1 above, it should be noted that the parameters
of each group distribution $f_j(\cdot|\theta_j),\ j=1,\ldots,K$ in the mixture
are independent given $\mathbf{z}$, i.e., $\theta_j$ and $\theta_{j\prime}$ (with
$j\neq j\prime$) are independent given $\mathbf{z}$. Hence, if each $\theta_j,\
j=1,\ldots,K$ is comprised of a single parameter (as in the mixture of Poisson
distributions or Gaussian distributions with known precision), then this will
hold. If this is not the case, then extra constraints may be required
to guarantee independence of the parameters in $\theta_j$, at least approximately.

For example, consider that $f_j(\cdot|\theta_j)$ is an Gaussian distribution
with mean $\mu_j$ and precision $\tau_j$. If the prior on $\tau_j$ is a Gamma
distribution with parameters $a$ and $b$, and the prior on $\mu_j|\tau_j$ is a
Gaussian distribution with mean $\mu$ and precision $\tau_j$, then the
full conditionals are given by

\begin{eqnarray}
\pi(\mu_j|\tau_j,\mathbf{z}, \mathbf{y}) \propto N\left(\frac{\mu+s_j}{1+n_j}, \tau_j(1 + n_j)\right)\nonumber\\
\pi(\tau_j|\mu_j, \mathbf{z}, \mathbf{y}) \propto Ga\left(a + \frac{n_j+1}{2},
  b + 0.5(\mu_j - \mu)^2 + 0.5 ss_j\right)
\label{eq:Gaussian-full-cond}
\end{eqnarray}
\noindent
In the previous expressions, $n_j$ is the number of observations assigned to
component $j$, $s_j$ the sum of their values and $ss_j$ is the sum of squares
of the values around mean $\mu_j$.

In this mixture of Gaussian distributions,
mean $\mu_j$ and precision $\tau_j$ are not conditionally independent. However, if the number of observations assigned to component $j$ in the mixture, say $n_j$, is large, then both parameters will be approximately independent
as data will be driving the values of the parameters in the full conditionals,
as seen in Equation~(\ref{eq:Gaussian-full-cond}).

The approximation in step 2 will be valid when the full conditional is
concentrated about its mode, i.e., it is (approximately) centered at the mode
and with a very small variance. This will often happen when data assigned to
each group in the mixture is large. Note that in the case of the mixture of
Gaussian distributions, it is clear from the full conditionals in
Equation~(\ref{eq:Gaussian-full-cond}) that their respective variances will
tend to zero as the number of observations in that group increases.

For a mixture of Poisson with $f_j(\cdot|\theta_j)$ a Poisson with mean
$\mu_j$ and a Gamma prior (i.e, $\pi(\mu_j) \sim Ga(a, b)$), the full
conditional distribution is

\begin{equation}
\pi(\mu_j|\mathbf{z}, \mathbf{y}) \propto Ga(a + s_j, b + n_j) 
\label{eq:Poisson-full-cond}
\end{equation}
\noindent
Here, $n_j$ is the number of observations assigned to component $j$ and $s_j$
the sum of their values.

Hence, in the Poisson case,
with full conditionals for means given in
Equation~(\ref{eq:Poisson-full-cond}), it is also clear that the variance will
tend to zero as the number of observation in that group increases.
Also, note that here $s_j$ will be approximately $\lambda_j n_j$ (under the
right model) and the variance of the full conditional will then be
$(a + \lambda_j n_j) / (b + n_j)^2 \simeq (\lambda_j n_j) / n_j^2 = \lambda_j / n_j$, which will tend to zero for a large $n_j$ under a vague prior (i.e., small values of $a$ and $b$ as compared to $n_j$).

\newpage 

\section{Galaxies dataset}
\label{app:galaxies}

Table~\ref{tab:galaxies-hc} shows a summary of the estimates obtained when
fitting mixture models with a half-Cauchy prior on the standard deviation to the
Galaxies datset. Similarly, Table~\ref{tab:galaxies-un} shows similar results
for mixture models with an improper uniform prior on the standard deviation.

\begin{table}[pos=h]
\centering
\begin{scriptsize}
\caption{Summary of results for the different mixture models fit on the Galaxies dataset.
Symbol $^*$ means that
model $\mathcal{M}_1$ has entirely been computed with INLA as it has a single component. $\hat\pi(\mathcal{M}_j|\mathbf{y})$ is the estimate of the posterior
probability of the models, which is the same regardless of the approximation
to the marginal likelihood used. Parameter estimates are summarized using
posterior mean and standard deviation (between parentheses).}
\label{tab:galaxies-hc}
\setlength{\tabcolsep}{0.5em}
\begin{tabular}{|c|c|c|c|c|c|c|c|c|}
Model & $\log(\hat\pi_S(\mathbf{y}))$ & $\log(\hat\pi_G(\mathbf{y}))$ & $\hat\pi(\mathcal{M}_j|\mathbf{y})$ &
  \multicolumn{5}{c}{Parameter estimates}\\
 & $\log(\hat\pi_M(\mathbf{y}))$ & & & Param. & Comp. 1 & Comp. 2 & Comp. 3 & Comp. 4\\
\hline
$\mathcal{M}_1$ &  -252.11$^*$ &  -252.11$^*$ &  0.00 & $\bm\mu$ & 20.82 (0.51) & -- & -- & -- \\
 & & & & $\tau$ &  0.05 (0.008) & & & \\
$\mathcal{M}_2$ & -247.87 & -248.94 & 0.00 & $\bm\mu$ & 9.85 (1.27) &  21.87 (0.36) & -- & -- \\
 & & & & $\mathbf{w}$ & 0.09 (0.011) & 0.91 (0.011) & -- & -- \\
 & & & & $\tau$ &  0.11 (0.018) & & & \\
$\mathcal{M}_3$ & -235.99 & -237.08 & 1.00 & $\bm\mu$  & 9.74 (0.81) & 21.40 (0.25) &
   32.91 (1.26) & --\\
 & & & & $\mathbf{w}$ & 0.09 (0.002) & 0.87 (0.004) & 0.04 (0.003) & -- \\
 & & & & $\tau$ &  0.23 (0.036) & & & \\
$\mathcal{M}_4$ & -243.04 & -245.23 & 0.00 &  $\bm\mu$  &  9.71 (0.53) & 20.05 (0.30) &  23.48 (0.37) &  33.02 (0.80) \\
 & & & & $\mathbf{w}$ & 0.09 (0.000) & 0.53 (0.049) & 0.34 (0.049) & 0.04 (0.000) \\
 & & & & $\tau$ &  0.54 (0.120) & & & \\
\end{tabular}
\end{scriptsize}
\end{table}

\begin{table}[pos=h]
\centering
\caption{Summary of results for the different mixture models fit on the Galaxies dataset.
Symbol $^*$ means that
model $\mathcal{M}_1$ has entirely been computed with INLA as it has a single component. 
Parameter estimates are summarized using
posterior mean and standard deviation (between parentheses).}
\label{tab:galaxies-un}
\setlength{\tabcolsep}{0.5em}
\begin{scriptsize}
\begin{tabular}{|c|c|c|c|c|c|c|c|c|c|}
Model & $\log(\hat\pi_S(\mathbf{y}))$ & $\log(\hat\pi_G(\mathbf{y}))$ &
$\hat\pi_S(\mathcal{M}_j|\mathbf{y})$ & &
  \multicolumn{5}{|c|}{Parameter estimates}\\
 & $\log(\hat\pi_M(\mathbf{y}))$ & &  $\hat\pi_M(\mathcal{M}_j|\mathbf{y})$ &
 $\hat\pi_G(\mathcal{M}_j|\mathbf{y})$& Param. & Comp. 1 & Comp. 2 & Comp. 3 & Comp. 4\\
\hline
$\mathcal{M}_1$ &  -244.79$^*$ &  -244.79$^*$ &  0.00 & 0.00 & $\bm\mu$& 20.82 (0.51) & -- & -- & --\\
  & & & &  & $\tau$ & \multicolumn{4}{c|}{0.05 (0.008)} \\
$\mathcal{M}_2$ & -241.35 & -242.42 & 0.00  & 0.00 & $\bm\mu$ & 9.99 (1.69) &  21.87 (0.39) & -- & -- \\
  & & & & & $\mathbf{w}$ & 0.10 (0.066) & 0.90 (0.066) & -- & -- \\
  & & & & & $\tau$ & \multicolumn{4}{c|}{0.10 (0.018)} \\
$\mathcal{M}_3$ & -230.22 & -231.30 & 0.86 & 0.72 & $\bm\mu$ & 9.74 (0.83) & 21.40 (0.25) &
   32.91 (1.28) & -- \\
  & & & & & $\mathbf{w}$ & 0.09 (0.002) & 0.87 (0.004) & 0.04 (0.003) & -- \\
  & & & & & $\tau$ & \multicolumn{4}{c|}{0.22 (0.036)} \\
$\mathcal{M}_4$ & -232.05 & -232.26 & 0.14 &  0.28 & $\bm\mu$ & 9.58 (3.3) & 20.12 (0.35) &  23.33 (0.69) &  33.01 (0.88)\\
  & & & & & $\mathbf{w}$ & 0.08 (0.022) & 0.50 (0.132) & 0.37 (0.152) & 0.04 (0.000) \\
  & & & & & $\tau$ & \multicolumn{4}{c|}{0.48 (0.139)}  \\
\end{tabular}
\end{scriptsize}
\end{table}



\end{document}